\documentclass{PoS}
\usepackage{natbib}
\bibpunct{(}{)}{;}{a}{}{,}

\title{Studies of Relativistic Jets in Active Galactic Nuclei with SKA}

\ShortTitle{Studies of Relativistic Jets in Active Galactic Nuclei with SKA}

\author{\speaker{Iv\'an Agudo}$^{1}$\thanks{On behalf of the SKA Cosmic Magnetism Working Group}, 
             Markus B\"ottcher$^2$, 
             Heino Falcke$^{3}$, 
             Markos Georganopoulos$^{4}$, 
             Gabriele Ghisellini$^{5}$, 
             Gabriele Giovannini$^{6}$, 
             Marcello Giroletti$^{7}$, 
            Jose L. G\'omez$^{8}$, 
             Leonid Gurvits$^{1,9}$, 
              Robert Laing$^{10}$, 
             Matthew Lister$^{11}$, 
             Jose-Mar\'ia Mart\'i$^{12}$, 
             Eileen T. Meyer$^{13}$, 
             Yosuke Mizuno$^{14}$, 
             Shane O'Sullivan$^{15}$, 
             Paolo Padovani$^{10}$, 
             Zsolt Paragi$^{1}$, 
             Manel Perucho$^{12}$, 
             Dominik Schleicher$^{16}$, 
             {\L}ukasz Stawarz$^{17}$, 
             Nektarios Vlahakis$^{18}$,
             and John Wardle$^{19}$\\ 

             {$^1$}Joint Institute for VLBI in Europe; 
              {$^2$}North-West University;
              {$^3$}Radboud University Nijmegen; 
              {$^4$}University of Maryland;
              {$^5$}INAF - Osservatorio Astronomico di Brera; 
              {$^6$}Universit\`a di Bologna; 
              {$^7$}INAF - Istituto di Radioastronomia; 
              {$^8$}Instituto de Astrof\'isica de Andaluc\'ia - CSIC; 
              {$^9$}Delf University of Technology;               
              {$^{10}$}European Southern Observatory; 
              {$^{11}$}Purdue University; 
              {$^{12}$}Universitat de Val\`encia; 
              {$^{13}$}Space Telescope Science Institute;
              {$^{14}$}National Tsing-Hua University; 
              {$^{15}$}The University of Sydney; 
              {$^{16}$}Georg-August-Universit\"at G\"ottingen; 
              {$^{17}$}Institute of Space and Astronautical Science --JAXA; 
              {$^{18}$}University of Athens;
              {$^{19}$}Brandeis University
  
              E-mail: \email{agudo@jive.nl}}

\abstract{ Relativistic jets in active galactic nuclei (AGN) are among
  the most powerful astrophysical objects discovered to date.  Indeed,
  jetted AGN studies have been considered a prominent science case for
  SKA, and were included in several different chapters of the previous
  SKA Science Book (Carilli \& Rawlings 2004).  Most of the
  fundamental questions about the physics of relativistic jets still
  remain unanswered, and await high-sensitivity radio instruments such
  as SKA to solve them.  These questions will be addressed specially
  through analysis of the massive data sets arising from the deep,
  all-sky surveys (both total and polarimetric flux) from SKA1.
  Wide-field very-long-baseline-interferometric survey observations
  involving SKA1 will serve as a unique tool for distinguishing
  between extragalactic relativistic jets and star forming galaxies
  via brightness temperature measurements.  Subsequent SKA1 studies of
  relativistic jets at different resolutions will allow for
  unprecedented cosmological studies of AGN jets up to the epoch of
  re-ionization, enabling detailed characterization of the jet
  composition, magnetic field, particle populations, and plasma
  properties on all scales. SKA will enable us to study the dependence
  of jet power and star formation on other properties of the AGN
  system.  SKA1 will enable such studies for large samples of jets,
  while VLBI observations involving SKA1 will provide the sensitivity
  for pc-scale imaging, and SKA2 (with its extraordinary
  sensitivity and dynamic range) will allow us for the first time to
  resolve and model the weakest radio structures in the most powerful
  radio-loud AGN.  }

\FullConference{
Advancing Astrophysics with the Square Kilometre Array\\
June 8-13, 2014\\
Giardini Naxos, Italy}

\newcommand {\apgt} {\ {\raise-.5ex\hbox{$\buildrel>\over\sim$}}\ }
\newcommand {\aplt} {\ {\raise-.5ex\hbox{$\buildrel<\over\sim$}}\ }

\newcommand\nar{New Astronomy Reviews}

\newcommand\apj{ApJ} 
\newcommand\apjl{ApJL}

\begin{document}

\section{Introduction}
\label{intr}
Relativistic jets in AGN are among the most powerful astrophysical
objects so far discovered.  Their relativistic nature causes them to
emit abundant and extremely time-variable radiation in all spectral
ranges from radio wavelengths to gamma-rays, whereas bulk relativistic
motion leads to strong Doppler boosting effects, making them
detectable at extreme cosmological distances.  In AGN, relativistic
plasma jets are thought to be formed as the result of accretion onto
super-massive black-holes (SMBH) in the presence of rotating accretion
disks and co-rotating magnetic fields.  The study of the details
involved in the jet formation process, which are still largely
unknown, makes AGN jets a powerful tool to probe the environment of
objects in extremely compact matter-states, and the physics of
high-energy plasmas and their magnetic fields.

During recent years, a drastic improvement in our understanding of
relativistic jets has been enabled by advances in supercomputing
facilities for state-of-the-art numerical simulations, as well as
deep, high-resolution radio, optical-IR and X-ray imaging and the
advent of comprehensive monitoring programs covering most accessible
spectral ranges from radio to gamma-rays.  However, most of the
fundamental questions about the physics of relativistic jets still
remain unanswered, and await high sensitivity radio instruments such
as SKA to solve them.  These questions include: a) why are jets
produced efficiently only in some systems, and what is the relation
between jet power and the properties of the black hole and accretion
system?  b) what influence do magnetic fields have on jet formation,
collimation, and maintenance up to distances $>100$\,kpc?  c) what is
the actual plasma composition at different jet scales and how does it
evolve down the jet?  d) how does the particle acceleration mechanism
make the jet an efficient emitter on scales exceeding the size of the
host galaxy? and d) how does the feedback between the jet and the
(inter-)galactic medium influence the evolution of galaxies and
clusters, and how do AGN jets and their central black holes evolve on
cosmological time scales up to $z\sim10$?

These questions will perhaps be addressed first by analysis of the
massive data sets arising from deep all-sky surveys with SKA1,
including both total and polarimetric flux.  In particular, SKA1
studies of relativistic jets at different resolutions will allow, for
the first time: i) studies (including imaging) of AGN jets throughout
cosmic time, back to the epoch of re-ionization; ii) robust
jet composition studies through unparalleled circular
polarization (CP) sensitivity/purity; iii) an understanding of the
origin of intrinsic differences in jet power in AGN via imaging of
thousands of radio-weak AGN; iv) a characterization of the
three-dimensional distributions of flow parameters such as velocity,
proper emissivity, and magnetic field structure (through total flux
and polarization imaging), that lead to estimates of key physical
quantities such as mass and energy fluxes and entrainment rates; v) a
probe of the magnetoionic environments of the jets (perhaps including
their confining fields) through deep full Stokes (and Faraday rotation
measure) imaging at all scales along and across a large number of
objects; and vi) an understanding of the interplay between the jet and
the ambient medium, both in the local universe and along cosmic time
up to $z\sim10$.

In \S~2--10, we describe the current status of the main research areas
involving jetted AGN, and the requirements necessary for SKA1 to
advance in these fields.  In \S~11 and 12 we summarize the general
need for future jetted AGN studies with SKA1, and we discuss the
prospects for the SKA1 early science phase, as well as for
SKA2 on AGN jet investigations.

\section{Jet Formation, Collimation, and Evolution}
\label{form}
\begin{figure}
    \centering
    \includegraphics[width=0.65\textwidth]{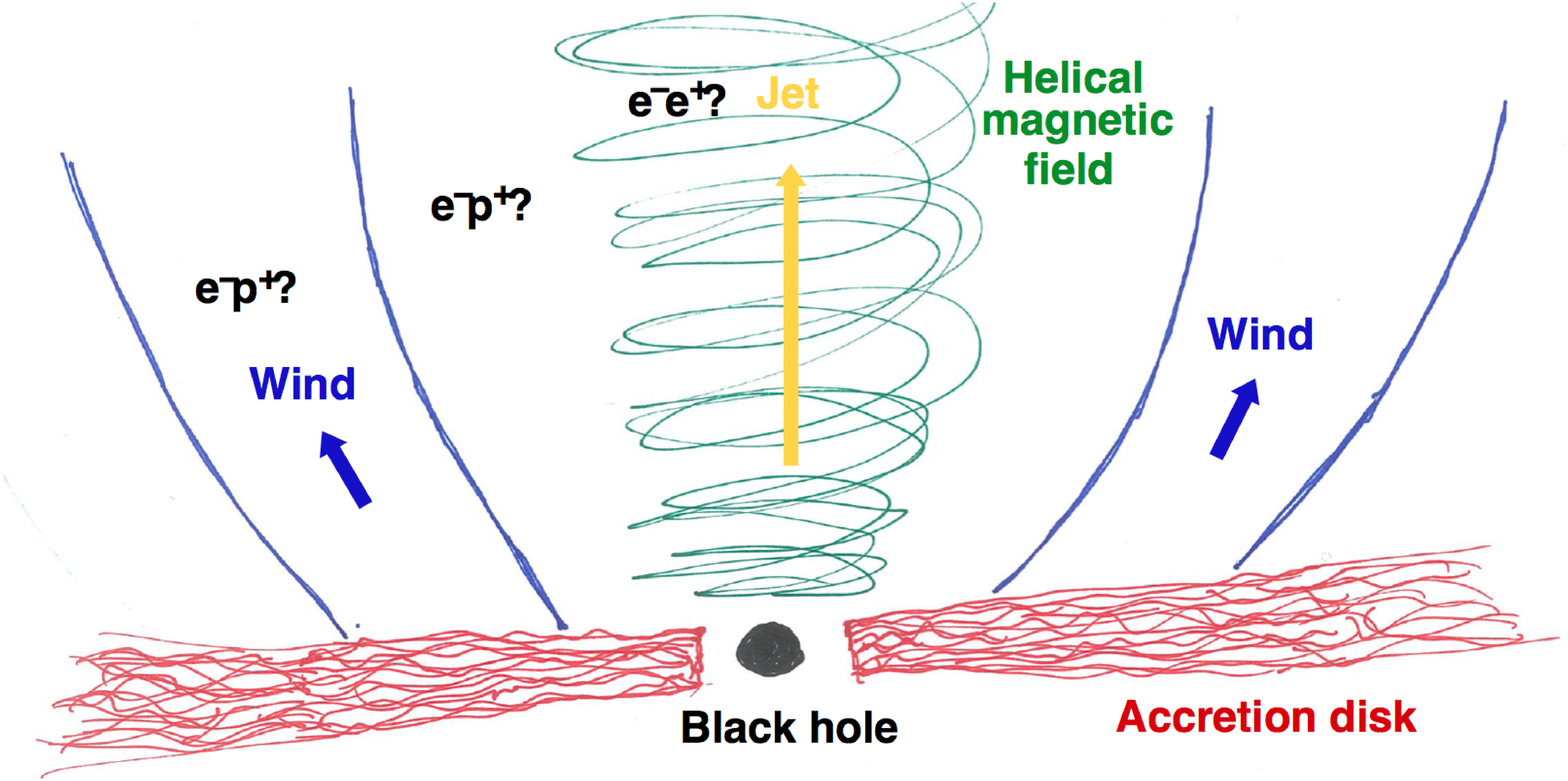}
    \caption{Conceptual illustration of the jet formation and collimation scenario described in \S~2.}
    \label{sketch}
\end{figure}

Models proposed to explain the origin of relativistic jets in AGN
involve accretion, in the form of a disk, of interstellar mass and gas
from tidally disrupted stars onto a supermassive black hole.  There is
a general agreement that magnetic fields play a crucial role in this
process.  In the models of magnetically driven outflows, poloidal
magnetic fields anchored at the base of the accretion disk generate a
toroidal component and a subsequent poloidal electromagnetic flux of
energy (Poynting flux) that accelerates the magnetospheric plasma 
along the poloidal magnetic field lines~\citep[see][and
Fig.~\ref{sketch} for illustration]{BlandfordPaine1982,Li1992}. Energy
can also be extracted from rotating black holes with similar
efficiencies by the \cite{BlandfordZnajek1977} mechanism. The strength
of the magnetic field on the event horizon can be estimated to be of
the order of thousands of Gauss. How this magnetic field is built up
from the disk magnetic field is the subject of current
research~\citep{Tch11,McK12}. 

Since in any version of a magnetic central engine the accretion disk
is magnetized, we expect a wind to be driven from its surface (close
to the black hole) with a power comparable to the
Blandford-Znajek power.  If the mass loading of disk winds (mainly
formed by electron-proton plasma) were too high to produce
ultrarelativistic terminal speeds, both mechanisms would need to
operate at the same time, producing stratified jets in both composition
(outer electron-proton wind, inner electron-positron jet) and speed.
On the other hand, the Blandford-Znajek mechanism does not provide any
collimation.  Hence the central engine must provide an additional
ingredient to confine the outflow until it reaches the superfast
magnetosonic regime. Multi-wavelength full-polarization (4-Stokes)
studies with SKA1 at all resolutions, down to the milli-arcsecond
(mas) scale with SKA1-VLBI \citep{Par14}, including CP observations
with precision up to $\sim0.01$\,\%, are ideal to test the above
mentioned jet/disc-wind scenario, and therefore of the role of such
winds in the collimation of jets (see \S~\ref{circ}).  Furthermore,
full-polarization measurements would allow us to determine the
fraction of electron-positron pairs, the shape of the low energy
distribution of the relativistic emitting particles and the presence
of a dynamically dominant proton component, helping to constrain the
initial magnetization of the wind, a very important parameter in the
magnetic theory of AGN jets determining the asymptotic speed of the flow.

Theoretical and observational considerations suggest that jets must
remain Poynting dominated along the acceleration phase.  The success
of magnetic acceleration \citep{VlahakisKoenigl2004} has been
confirmed in time-dependent numerical simulations
\citep{Komissarov2007}.  In the case of parabolic jets, the
acceleration continues until the jet becomes particle-dominated, at
sub-parsec scales (hundreds to thousands of black hole gravitational
radii).  To stablish the exact location of the site of this conversion
and its possible observational signature, in particular its position
in relation to the radio core, is a subject of current active research
for multi-spectral-range observational campaigns where high-frequency
polarimetric single-dish and VLBI observations are key ingredients.
With its superb sensitivity, SKA1-MID in Band 5 (both as a VLBI
station or not) will allow these kinds of studies for a large number
of sources seen from the southern hemisphere, where a growing number
of panchromatic facilities are expected to operate with SKA.  These
studies are particularly interesting for nearby sources such as
Centaurus\,A and M\,87, where the highest linear resolutions will be
achieved to resolve the jets both across and along their axis, from
the regions close to the SMBH to those where the Poynting flux
converts to particle dominated plasma.

\section{Particle Acceleration in Jets}
\label{acc}
The ambiguity of the location where Poynting-flux dominated outflows
are converted into kinetic dominated jets reflects a general lack of
understanding of the particle acceleration processes which are
responsible for maintaining power-law distributions of radiating
ultra-relativistic electrons over decades of length scales along the
jet. These scales range from the sub-pc to pc scale, where the
high-energy emission from blazars is likely produced, to the kpc scale
where the jets terminate. In the case of low plasma magnetization,
diffusive acceleration taking place at the fronts of strong
hydrodynamical shocks \citep[see e.g.,][and references therein]{sb12}
could be considered as the most efficient way of energizing jet
particles, leading naturally to the formation of power-law spectra.
In the case of a high magnetization, on the other hand, relativistic
magnetic reconnection \citep[e.g.][]{sir14} and stochastic
interactions of particles with magnetic turbulence \citep[][and
  references therein]{sta08} are expected to play a major role.
Although the relativistic nature of AGN jets constitutes a difficulty
in this context (the aforementioned particle acceleration processes
are well understood only in a non-relativistic regime), the slope and
normalization of the resulting non-thermal particle spectra depend
strongly on the underlying physical conditions in all of these three
cases.  This offers the opportunity to exploit high-quality
observational data to determine the underlying non-thermal particle
distributions, which can be used to constrain the acceleration
processes at work.

Good-quality radio data
provided by SKA1 has the potential to reveal the spectral shapes of
freshly accelerated jet electrons.  In the case of the extended
arcsecond-scale jets and jet cocoons (lobes), such studies are
expected to be free of any absorption effects even at low frequencies
($<1-2$\,GHz).  The superior SKA1 sensitivity in full
polarimetric mode (at $\apgt1$\,GHz), combined with
simultaneous SKA1-VLBI observations, will enable
spatially resolved spectral studies of the radio emission of the
innermost regions of AGN jets out to large redshifts, ($z \sim 5$ --
10).  Additional constraints on the acceleration sites of the
highest-energy electrons and their radiation environment, based on
co-ordinated multi-wavelength observations at high radio frequencies
(see previous section), will be facilitated by
the location of the SKA1-MID in South Africa, in close
vicinity to world-class optical (SALT) and very-high-energy $\gamma$-ray
(H.E.S.S. and possibly CTA-South) observatories, in
combination with satellite-based X-ray and $\gamma$-ray (Fermi)
observations.

\section{Jet Composition through Circular Polarization Studies}
\label{circ}
Circular polarization observations (Stokes $V$) of relativistic jets
enable us to measure quantities that are not accessible by observation
of the first three Stokes parameters ($I$, $Q$, $U$, representing
total intensity and linear polarization, LP).  These include direct
measurement of the magnetic field strength, as well as the
vector-ordered part of the magnetic field, and hence the true magnetic
flux carried by the jet (a conserved quantity that should be equal to
the magnetic flux at the central engine). CP will also enable
measurements of the composition of the jet (i.e.~the e$^+$-p$^+$
ratio) and the low-energy cut-off of the relativistic electron energy
distribution \citep[see][where all these physical quantities were
  estimated for the quasar 3C~279]{Homan2009}.  These quantities are
fundamental to understanding the physical properties of relativistic
jets and in particular how they are launched (see \S~1 and 2, and
Fig.~\ref{sketch}), and also for calculating the energy carried by the
jet and hence for studying AGN feedback.  These have long been open
questions in AGN physics.

CP is difficult to study in AGN because it is typically weak
($\sim$0.1--1\% of Stokes $I$) and highly variable.  The high
sensitivity and large instantaneous bandwidths of the SKA1 make it the
ideal instrument for measuring CP (and its frequency dependence up to
Band 5 of SKA1-MID), as long as the SKA retains excellent polarization
purity at all frequencies.  As very few circularly polarized AGN are
currently known ($<100$), a large-area CP-sensitive SKA1 survey across
all SKA1 bands from 1 to 20~GHz will revolutionize this field of study
\citep[see][for an example of what can be achieved with broadband CP
  observations]{Osu13}.  
After finding the CP-brighter sources, SKA1-VLBI is required for
follow-up observations to resolve the CP emission region and extract
the key jet physical parameters (power, composition, magnetic flux).

CP measurements are unique in their ability to provide a direct
measurement of the jet magnetic field strength, if the field vector
ordering can be determined.  One situation where there must be a
vector-ordered field is if the jet carries an electric current, as in
several of the electromagnetic mechanisms for launching jets.  Indeed,
when combined with LP and Faraday rotation measurements, several
fundamental properties of the jet can be measured using CP.  For
example, when a transverse CP gradient and an RM gradient are both
detected, they should have the same sign, each confirming the other,
allowing direct measurement of the strength of the toroidal component
of the magnetic field and hence the magnitude of the jet current, as
well as the low energy cutoff of the electron energy distribution
(from the RM-CP ratio).  Since we are trying to detect very small
fractional polarizations, our detected sources will be bright in
Stokes $I$.  Thus, as long as SKA1 receivers have excellent
polarization purity ($\sim0.1$\,\% and $\sim0.01$\,\% for LP and CP,
respectively), then even a shallow large-area early-science survey
will dramatically improve our knowledge of CP in AGN jets.  Sources
detected in CP could then be followed up with
SKA-VLBI to produce groundbreaking advances in our knowledge of
magnetic fields and particle populations in jets.

\section{From the Most Powerful to the Weakest Jets}
\label{rwjets}

\subsection{The Radio-loud/Radio-quiet AGN Issue}
\label{AGNdichotomy}
\begin{figure}
    \centering
    \includegraphics[width=0.4\textwidth]{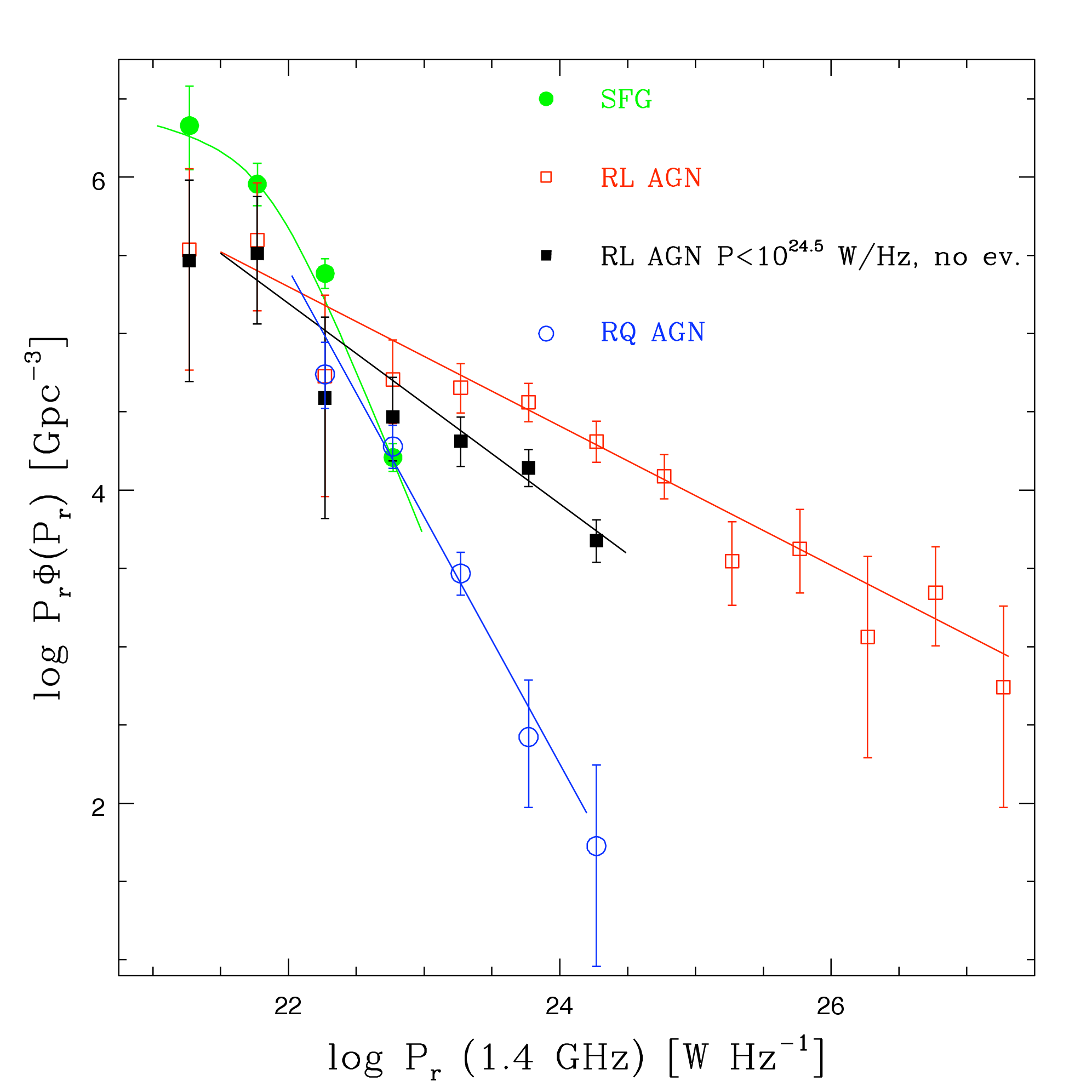}
    \caption{Local 1.4\,GHz differential luminosity functions
      for star forming galaxies and AGNs measured from the \emph{Chandra}
      Deep Field South VLA survey. Reproduced from
      \citet{Padovani_etal11}.}
    \label{pad}
\end{figure}

AGN come mainly in two classes: radio-loud (RL) AGN, emitting most of
their energy over the entire electromagnetic spectrum non-thermally
through powerful relativistic jets, and radio-quiet (RQ) AGN, whose
multi-wavelength emission is dominated by thermal emission, directly
or indirectly related to the accretion disk\footnote{See
  \citet{Bonzini13} for a recent definition of radio-quiet and
  radio-loud AGN}.  The mechanism responsible for radio emission in RQ
AGN has been a matter of debate for the past fifty years.
Alternatives have included a scaled down version of the RL AGN
mechanism \citep[e.g.,][]{Ulvestad_etal05}, star formation
\citep{Sopp_91}, and many others.  Very recent results on the
evolution and luminosity function of very faint radio sources, based
on the Chandra Deep Field South samples including a few hundred sources down to $\sim 6
-10~\mu$Jy rms, suggest close ties between star formation and radio
emission in RQ AGN, at least at $z \sim 1.5 - 2$
\citep[][Fig.~\ref{pad}, and Padovani et al. in
  prep.]{Padovani_etal11}.  This is further confirmed by the
comparison of the star formation rates derived from the far-IR and
radio luminosities, assuming that all the radio emission is due to
star formation.  For RQ AGN and star forming galaxies the two star formation rate
estimates are fully consistent, while for RL AGN the agreement is poor
due to the large contribution of the relativistic jet to their radio
luminosity (Bonzini et al., submitted).  SKA will contribute in two
different ways to examine this problem, i) by using the data from the
full-polarimetric commensal wide and deep surveys that are being
planned for SKA1 in Band 3, which will provide larger samples of RQ and
RL AGN covering a wide range of redshifts and powers, and ii) by
carrying out polarimetric, wide-field SKA1-VLBI survey
follow-up observations in Bands 3 and 5 of the brighter sources
($\apgt50\,\mu$Jy) to distinguish between jet and star formation
emission (see \S~\ref{ident}).

\subsection{Jet Power Along the Radio-loud Classification}
The more powerful radio-loud AGN jets are traditionally divided into
Fanaroff-Riley classes FRI and FRII, with the latter having highly
collimated jets that terminate in a bright hotspot, indicative of
high-Mach number, relativistic flows \citep{1974MNRAS.167P..31F}.
Since the pioneering work of \citet{rawlings91} on radio-loud AGN,
that showed that the jet power correlated with the narrow line
luminosity, we have reached the firm conclusion that the jet power
correlates with the accretion rate.  The present large sample of
spectroscopically observed radio--loud objects (SDSS), and blazars
detected in the $\gamma$--ray band reinforced this conclusion
\citep{sbarrato12a}, and led to the idea that low power radio--loud
AGN are associated with radiatively inefficient disks, emitting a
small amount of ionizing photons, thus explaining why the broad
emission lines are so intrinsically weak in these objects
\citep{sbarrato12b}.  Of course, black hole mass and black hole spin
are also thought to play key roles
\citep[e.g.,][]{BlandfordZnajek1977,2011ApJ...727...39M} by
controlling the accretion rate, whereas the external environment also
has a strong influence on jet luminosity, as witnessed by `hybrid'
AGN, which exhibit FR I and FR II morphology jets on either side of
the active nucleus \citep{2000A&A...363..507G}.

\begin{figure}
    \centering
    \includegraphics[width=0.45\textwidth]{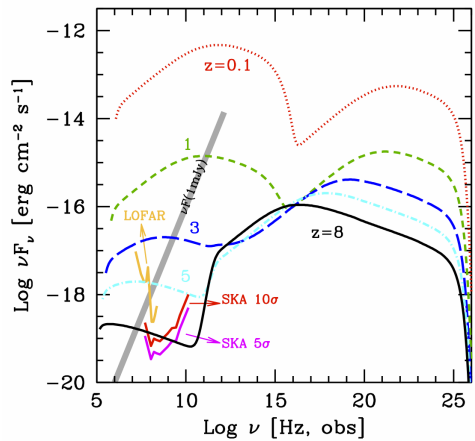}
    \caption{Spectral energy distribution
      of a typical radio loud quasar at different redshifts as seen in
      the observer's frame. As redshift increases, the radio spectrum
      gets progressively more dimmed both because of the larger
      distance to the source and because of inverse Compton cooling of
      the particle population responsible for the radio emission. The
      red and pink continuous lines mark the $10\sigma$ and $5\sigma$
      limits of SKA1 (LOW and MID), respectively, for 30\,min
      integration time, whereas the thick grey line correspons to $F(\nu)=1\,mJy$. Reproduced from \citet{ghisellini14}.}
    \label{ghis}
\end{figure}

The weaker flavor of radio-loud AGN have a synchrotron emission
peaking in the far--UV and X--ray bands, and are radio--weak (but not
radio--quiet).  To study them at moderate redshifts ($z>0.4$), it is
crucial to measure the IR and optical cosmic background and to
constrain the cosmic magnetic field.  The large and deep commensal
surveys planned for SKA1 will be an extremely useful tool to help
studying weak radio AGN associated with TeV emitters observed with
CTA \citep[see ][]{Gir14}, and to provide a better characterization of
the large scale properties of the cosmic magnetic field \citep{Joh14}.
Moreover, a significant fraction of Seyfert galaxies have radio
structures on scales of $\sim 1$ kpc consistent with weak jets being
decelerated by the host galaxy's ISM \citep{2006AJ....132..546G}.
High-sensitivity, high-resolution full polarimetric studies with SKA1
and SKA1-VLBI have the potential for imaging many of these sources, to
shed new light on the long standing problem of radio-loud/radio-quiet
AGN through high resolution imaging of nearby sources.

On the opposite power end, powerful jets are associated with high
rates of accretion \citep{ghisellini10}.  The most powerful objects
are therefore those jets associated with high accretion and large SMBH
masses.  Exploiting the relativistically enhanced jet emission, high
power blazars can be an important tool to explore the far Universe, to
find the youngest and most active SMBH, that can constrain (and
possibly challenge) current ideas of black hole formation, growth, and
feedback \citep[see e.g.,][]{volonteri11}.  For each detected blazar,
there must be hundreds of similar sources whose jet is pointing
elsewhere.  It is this that makes the search for high--redshift
blazars with large SMBH masses rewarding.  To this task, deep SKA1
surveys at Band 3 or lower frequencies will be crucial.  Not only
because fluxes weakens with distance, but because, at high $z$, the
cosmic background radiation (CMB, whose energy density $\propto
(1+z)^4$), makes the relativistic electrons emit more through the
inverse Compton scattering than through the synchrotron process.  As a
result, the fluxes of powerful blazar jets could be very weak at radio
wavelengths \citep[see][and Fig.~\ref{ghis}]{ghisellini14}, but
still detectable for the SKA1 below 3\,GHz for reasonably short
integrations ($\sim30$\,min) up to redshifts $z\sim8$ and perhaps
larger.  Deep follow-up Band 3 SKA1-VLBI observations of the first AGN
detected will enable studies of the jet morphologies at very high
redshifts, and therefore their dense environments.  This is also
required to discover how the CMB modifies the appearance of radio
lobes at high--$z$.  SKA1 can therefore revise the fraction of
radio-loud to radio-quiet AGNs at all redshifts, with a strong impact
not only on the radio--loud luminosity function and evolution, but,
in general, on the feedback on the host galaxy (through the so
called radio-mode).

\section{Detailed Estimates of Fundamental Jet Parameters}
\label{jetpar}
Deep, multifrequency, transverse-resolved radio observations are
crucial to the estimation of flow variables in jets.  This section
concentrates on two topics that are fundamental to jet formation and
propagation: velocity fields and magnetic-field topology.  One
powerful observational technique exploits the intrinsic symmetry of
relativistic jet formation.  The observed differences between
approaching and receding jets are then dominated by the effects of
relativistic aberration \citep{Aloy}, so by fitting parametrized
models to images of both jets in Stokes $I$, $Q$ and $U$, it is
possible to derive the jet inclination and the three-dimensional
variations of velocity, rest-frame emissivity and magnetic-field
ordering \citep{LB14}.  A second technique, which can use the same
observations, is to analyse the wavelength dependence of linear
polarization (and hence internal and external Faraday rotation) across
the observing band using RM synthesis.  These methods can be used on
any scale, given adequate sensitivity and angular resolution.
Structural changes on pc scales on timescales of days--years can be
monitored, allowing direct measurement of proper motions and the
variability of polarization and Faraday rotation structures.

\subsection{Velocity Field}
Low-luminosity (FR\,I) jets have been shown to decelerate on kpc
scales, developing transverse velocity gradients indicative of
boundary-layer entrainment \citep[see][]{LB14,Lai14}.  
Further progress requires modeling of
large samples, to average over intrinsic asymmetries, to explore the
dependence of deceleration on jet power and environment and to assess
the relative importance of mass injection by stars and boundary-layer
entrainment.  Resolutions in the range 0.1 -- 0.5 \,arcsec and rms
noise levels of 0.1 -- 0.5\,$\mu$Jy/beam at GHz frequencies are needed
to observe a sufficiently large number of sources ($\sim300$) with
SKA1-MID, Bands 3 -- 5.

The velocity fields of the narrower kpc-scale jets in powerful sources
are much less well determined. Here, a key question is whether they
have the very fast ($\Gamma \approx 5 - 10$) spines needed to generate
X-ray emission by inverse Compton scattering of CMB photons as well as
slower boundary layers.  To answer this question, much better
resolution and sensitivity are needed ($\approx$0.01\,arcsec,
$\approx$10\,nJy/beam and dynamic range $>10^7$:1; SKA2 Band 4 and
5).  See \citet{Lai14} for a more detailed use case.

\begin{figure}
    \centering
    \includegraphics[width=0.9\textwidth]{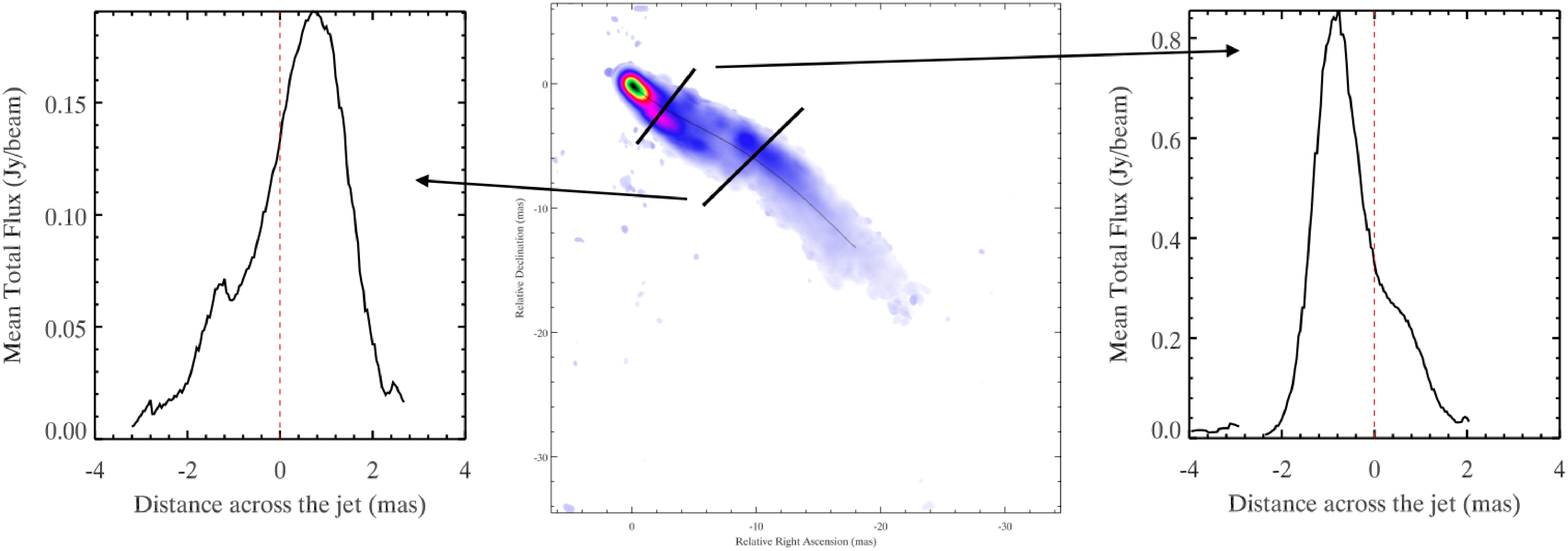}
    \caption{Mean total intensity map of 3C~273 obtained by combining
      48 epochs spanning 14 years of VLBA observations at 15 GHz
      \citep[including data from the MOJAVE program][]{MOJAVE}. Left and right
      panels show slices at 3.1 and 14 mas from the core. A black line
      extending from the core indicates the jet ridge-line.}
    \label{fig:3c273}
\end{figure}

The increased sensitivity of SKA1-VLBI would in principle allow the
technique to be extended to pc scales.  Stratification in the total
and linearly polarized emission across the jet width are expected in
case the synchrotron emitting jet is threaded by a helical magnetic
field \citep{Laing81,Aloy}.  In this case, variations in the velocity
field can be estimated by searching for changes in the emission
stratification along the jet.  Such variations have been observed in
the jet of 3C~273 -- for which there is evidence for the existence of
a helical magnetic field \citep{Asada} -- by combining data from 48
epochs spanning 14 years of VLBA observations at 15\,GHz
\citep[see][and Fig.~\ref{fig:3c273}]{Gomez12}.  
A similar jet behavior was already observed with 5\,GHz VLBI by \citet{Lob01}.
A change in the side
of the jet that is brighter suggests an acceleration of the emitting
plasma, in agreement with the observed acceleration of the pattern
velocity associated with superluminal components
\citep{Jorstad,Lister}.  The unprecedented sensitivity of SKA1-VLBI
will allow us to extend these studies to statistically significant
samples of AGN jets.

For sources in which both jets are visible it is also possible to
estimate the flow speed on scales where the pattern speeds of moving
components can also be measured.  If pattern speeds can be measured
for components emitted simultaneously in the approaching and receding
jets (as can be done for some microquasar jets), then additional
constraints can be derived.  M\,87 is a particularly important target,
as the hypothesis of gradual MHD acceleration on scales $\gg R_{\rm
  Schwarzschild}$ can be tested \citep[see e.g. ][]{M87} .  Note,
however, that the maximum baseline and observing frequency both need
to be high enough to ensure transverse resolution of the jets and
removal of any associated Faraday rotation \citep{Kovalev}, which
makes polarimetric VLBI involving SKA1-MID up to Band 5 the ideal
option.

If the velocity field, geometry and external environment of a jet are well 
characterized, then it is possible to apply conservation laws to derive the
 fluxes of energy, mass and momentum and the variation of entrainment 
 rate with distance from the AGN \citep{LB02}.

\subsection{Magnetic Field Topology}
Most models of jet formation require a vector-ordered (toroidal or
helical) field, but this has proved to be difficult to confirm
observationally.  The alternative of a random field made anisotropic
by compression (e.g.\ shocks) or shear often provides at least as good
a fit to observations of linear polarization.  There are two key
diagnostics.  Firstly, a helical field with a significant poloidal
component shows characteristic asymmetries in its transverse intensity
and polarization emission profiles \citep{Aloy}.  Secondly, if thermal
matter is present within the jet volume, then systematic transverse
Faraday rotation gradients are expected \citep{Laing81,Broderick}.
Both diagnostics are complicated by effects such as jet bending and
foreground Faraday rotation.  The best-observed cases on pc scales,
3C\,120 \citep{Gomez11} and 3C\,273 \citep{Asada} suggest a slow
foreground screen and jet-related Faraday rotation, respectively.
Deep and transverse-resolved observations of the pc-scale emission
from a large sample of straight jets are needed, to answer the
following questions: Are Faraday rotation gradients preferentially
orthogonal to the jet? Does RM synthesis evidence the
mixing of thermal and emitting plasma? What are the relative senses of
gradients in the approaching and receding jets? Are transverse
emission profiles consistent with the observed rotation? These
questions require SKA1-VLBI on Band 5.

The dominant field component on kpc scales in FR\,I jets is toroidal,
but it is unclear whether it is a vector-ordered remnant of the
original confining field.  If so, systematic and symmetrical gradients
in Faraday rotation across the jets are expected.  Isolating these
from confusing effects of Faraday rotation by the IGM
should be feasible for large samples of jets in sparse environments.

\section{Understanding Jet Emission at Sub-pc and Multi-kpc Scales}
Relativistic jets in AGN carry enormous amounts of kinetic energy out
from the central black hole into the kpc-scale galaxy and intercluster
medium.  Most of the energy dissipation in powerful blazars is through
the emission of GeV photons via inverse Compton scattering.  However,
we lack a clear understanding of where this dissipation takes place.
Is it very near to the black hole engine (sub-pc scale) or much
further out, beyond the regions dominated by broad-line emitting
clouds and the obscuring torus of the unified model?  As these scales
can differ by orders of magnitude, the answer gives important clues
for constraining jet formation and collimation models.

The dominant view during the {\sl EGRET} era in the 1990's was that
the gamma-ray emission is taking place within the sub-pc scale broad
line region \citep{sikora94}.  However, recent multi-wavelength
monitoring of a few bright sources from radio to the {\sl Fermi} GeV
band, strongly suggests that in at least some cases, the emission
takes place at distances of {\sl several pc} from the central engine.
The key evidence for this is VLBI observations
showing that (some) major $\gamma$-ray flares take place at the same
time with the ejection of superluminal components from the radio core,
located several pc downstream of the central engine
\citep[e.g.][]{marscher10}, or even co-spatial with VLBI components
downstream of the core \citep[e.g.][]{agudo11}.

The SKA1-MID observational setup will have suitable features to
monitor a substantial number of jets with the ideal sensitivity in
total intensity and polarization, angular resolution, and time
cadence: in particular Band 5 receivers in VLBI mode \citep{Par14}
will provide $\sim$ milliarcsecond resolution, which corresponds to
$\sim$ parsec scales for a broad redshift range.  The polarization
capabilities will be particularly relevant: current work suggest that
polarized features can help in the identification and tracking of jet
components and in the characterization of the magnetic field evolution
in relativistic jets.  Only a massive project based on SKA data,
perhaps employing similar observing methods and external multi-spectral range
infrastructures as other transient objects \citep[see][]{Fen14}, will
allow us to put these indications on firm grounds.

On much larger (kpc) scales, {\sl Chandra} showed that knots in the
jets of powerful quasars are strong X-ray emitters, with the X-ray
emission being a separate spectral component from the radio-IR-optical
one \citep[first seen in PKS 0637-752,][]{schwartz00,chartas00}.  The
standard interpretation is
that the X-rays are from inverse Compton scattering, assuming that the
radio emitting electron distribution extends down to electron Lorentz
factors $\gamma\sim10-100$ and that the jets are highly relativistic
at kpc scales \citep[bulk Lorentz factors
  $\Gamma\sim10-20$,][]{tavecchio00,celotti01}.  This mechanism
requires a jet power comparable to or higher than the Eddington
luminosity and jets that are very well aligned to the line of sight
and have lengths sometimes exceeding 1 Mpc.  An alternative
interpretation is that the X-ray emission is synchrotron coming from a
second population of electrons \citep[for a review see][]{harris06}.
This does not require highly relativistic speeds or extremely powerful
jets, but it does require that in situ particle acceleration reaches
electron energies of at least $\sim30$\,TeV.

We are faced, therefore with the choice of fast and powerful jets or
slow and multi-TeV particle accelerator jets.  Recently, it was shown
through {\sl Fermi} observations that in the case of the archetypical
quasar 3C 273 the inverse Compton model is ruled out \citep{meyer14},
favoring synchrotron from a second electron population as the X-ray
emission mechanism \citep[see also][for the case of PKS~1136$-$135
  studied through optical/UV polarimetry]{Car13}.  Multi-wavelength
observations of 3C 273 hint at the two electron populations not being
co-spatial \citep{jester05}.  Understanding if this is indeed the case
is imperative for understanding the physical conditions in the knots
of these powerful jets, and high-dynamic-range deep radio-images from
SKA1-MID (Bands 1--5) at resolutions up to 0.1 arcseconds (and
SKA-LOW) are essential for mapping the spectral energy distribution of
the extended jet as it extends out into the IGM.

\section{Feedback Between AGN Jets and their External Medium}
\label{feed}
In the last few decades, observations of
clusters of galaxies in the X-ray range revealed a lack of cool gas in the
centres of many galaxies, contrary to expectations \citep[see
  e.g.][and references therein]{Fabian12}.  This fact has not only important
implications regarding the star formation rates in the galaxy and the
growth of the central black hole \citep[e.g.][]{Cattaneo09}, but also
regarding cold dark matter (CDM) models \citep{Benson03}.  
The most accepted heating mechanisms proposed to stop
the cooling flows are related to galactic activity or shocks produced
during the merging of clusters.  Among the former, the observed
anti-correlation between the radio lobes formed by jets and the X-ray
emission from the cluster gas suggests that the buoyancy of the
under-dense cavities formed by the lobes could generate displacement
work on the ambient gas, and could be responsible for the heating.
Moreover, significant levels of metallicity have been recently
detected at considerable distances from the active galaxy, which is
possible only if an outflow has dragged those metals, produced from stars
in the galaxy \citep[e.g.][]{Werner11}.

Recent observations have shown that the lobes are surrounded
by shocks with low Mach numbers ($M \simeq 1 - 2$) in powerful
sources, i.e., they have not reached the buoyancy stage.  This would
imply faster evolution (pressure driven) and, accordingly, larger jet
powers than estimated.  Shocks may be
extremely important and efficient in the heating process of the
galactic and cluster gas because they can displace large amounts of
gas from the host galaxy, quenching or triggering star formation
\citep[e.g.][and references therein]{Perucho14}.  The
discrepancy between CDM models, which predict a power law distribution
of dark matter halos around galaxies, and the luminosity function of
bright galaxies, which declines exponentially, could be also related
to the feedback process \citep{Benson03}.  Recent semi-analytical work taking
into account AGN feedback, reproduces the bright end of the luminosity function
\citep[e.g.][]{Somerville08}.

It is clear that the dynamical impact of AGN jets within their host
galaxies and environment can be very important to their evolution.  
The way in which the galactic activity relates to heating and cooling flows, or affects the evolution
of the host galaxy, by changing the star formation rates and the growth of the central black hole, represents a fundamental problem in Astrophysics and Cosmology.
Old populations of
particles in the radio lobes formed by jets will certainly provide
deep insight into the interaction between radio-jets and their
environments.  These particles emit at low radio frequencies and make
SKA1-MID and SKA1-SUR (Bands 1 -- 3), as well as SKA1-LOW, ideal
instruments to detect them, and study how AGN jet activity impacts the
evolution of the host galaxy and intra-cluster medium via the
injection of particles and energy through relativistic jets
\citep{Har14}. Angular resolutions on the order of the arcsecond
($\sim10$\,arcseconds for SKA1-LOW), and sensitivities of the order of
the $\mu$Jy/beam in wide band continuum mode will be needed for these
studies.  The polarimetric capabilities of SKA will be relevant to
study the influence of magnetic fields on the interplay of large
scale-jets with their environment.

\section{AGN Jets Along Cosmic Time}
\label{cosm}
\subsection{The First AGN and their Supermassive Black Holes}
While AGN are common in the present-day Universe and SMBH with masses
larger than $10^9$~M$_\odot$ were still reported at $z>6$
\citep[e.g.][]{Fan06}, it is natural to ask when the first AGN and
their supermassive black holes started to form.  In a seminal
discussion by \citet{Rees78}, a number of potential pathways,
including single stars, stellar clusters and the direct collapse of
very massive gas clouds have been considered.  While the first stars
are expected to have typical masses ranging from a few $10$~M$_\odot$
to a few $100$~M$_\odot$, the formation of stellar mass black holes
from primordial stars seems only possible between $30-120$~M$_\odot$
and $180-1000$~M$_\odot$ \citep{Heger01}.  At least under the
assumption of Eddington accretion, it therefore seems difficult that
such stars could reach $10^9$~M$_\odot$ by $z\sim6-7$
\citep[e.g.][]{Shapiro05}.

More massive black hole seeds could be formed from the collapse of
dense stellar clusters. Such clusters could form via dust-induced
fragmentation in low-metallicity gas, with their collapse leading to
characteristic black hole masses of a few times $1000$~M$_\odot$
\citep{Devecchi12}.  The first simulations exploring the formation of
massive black holes from the direct collapse of massive gas clouds
have shown that strong radiation backgrounds are required to prevent
the fragmentation into low-mass objects \citep{Bromm03}.  Theoretical
suggestions indicated that masses of $10^4-10^5$~M$_\odot$ could be
reached from the low-angular momentum material in the first massive
halos at $z\sim10-15$.  Numerical simulations indeed indicate the
formation of massive self-gravitating disks in their centers, with
high accretion rates of $\sim1$~M$_\odot$~yr$^{-1}$.  The high
accretion rates imply the formation of very massive protostars with
cool atmospheres and weak radiative feedback, leading to a typical
mass scale of $10^5$~M$_\odot$ \citep{Schleicher13}.

These massive black holes are expected to lead to the formation of the
first high-redshift quasars, which may contribute to the epoch of
reionization.  The pathways to
observationally probe this population with SKA were outlined by
\citet{Falcke04}.  The compact radio cores of quasars with black hole
masses of $10^7$~M$_\odot$ should be visible even at $z>10$.  Due to
the dense environments in the first galaxies, \citet{Falcke04} predict
that the black hole jet can potentially not break out from the galaxy,
therefore leading to the formation of GHz-Peaked-Spectrum (GPS)
sources \citep[see also][]{Tay14,Afo14}.  Early-Universe GPS sources
are expected to be detected at low frequencies (100 -- 600\,MHz) at
sensitivities of 100 $\mu$Jy and better.  This makes wide bandwidth
surveys on SKA1-LOW, and SKA1-MID/-SUR at Band 1, the best suited
options for the detection of these sources.  Also, HI spectroscopy via
the $21$~cm line, potentially allows for the detection of hundreds to
thousands of bright QSOs via their HII regions at $z\sim8-10$; see
\citet{Wyi14} for an extended discussion.  Later, SKA1-VLBI involving
SKA-MID and SKA-SUR at the lowest practical frequencies (i.e. L Band,
to maximize UV coverage with additional existing VLBI stations) are
probably the right selection to confirm the compact nature of the GPS
sources down to scales smaller than a few tens of milliarcsecond.

\subsection{Cosmological Studies of Radio-loud AGN. From $z\sim0-10$}
Radio observations are the best-suited means to probe the evolution of
radio jets in SMBH over cosmic time, as obscuration has a very low
effect and largely unbiased surveys can thus be pursued
\citep{Falcke04}.  The latter allows cosmological studies of black
hole growth over a large range of black hole masses and cosmological
redshifts.  The NRAO VLA Sky Survey \citep[NVSS,][]{Condon98} has
detected $1.8$ million sources in total intensity, with $14\%$ of them
showing a clear polarization signal of at least $3\sigma$ (the majority of these being radio-loud AGN).  
While typical source counts have been restricted to flux density
sensitivities of $\sim10\mu$Jy in total intensity
\citep[e.g.][]{Windhorst03}, bright sources above $100$~mJy typically
correspond to the radio-loud FRII sources, while a gradual transition
towards FRI sources occurs around $30$~mJy.  At $\sim1$~mJy, an increasing fraction of radio sources is dominated by star
formation rather than an active galactic nucleus (see
\S~\ref{AGNdichotomy}), even though a contribution from a radio-quiet
AGN may still be present.

An analysis of the polarization data of NVSS shows that the fractional
polarization increases with decreasing flux density \citep{Mesa02}.
This trend has been confirmed in subsequent studies and surveys \citep[see][and references therein]{Banfield11},
including data from the Dominion Radio Astrophysical Observatory Deep
Field polarization study and radio polarimetry of the ELAIS N1 Field.
This intriguing anti-correlation clearly suggests a change in the
magnetic field structure of the observed sources, even though the
physical origin of this behaviour is not yet understood.
To investigate the potential origin of the polarization, \citet{Shi10}
studied a sample of highly polarized objects.  In their sample,
sources with more than $30\%$ polarization were contained in
elliptical galaxies with $1.4$~GHz luminosities in the range of
$10^{23}-10^{24}$~W~Hz$^{-1}$.  As they reported no dependence on
optical morphology, redshift, linear size and radio power, the high
polarization seems likely a result of intrinsic properties of the
source \citep[see also][]{Stil14}. 

The deep total flux and polarimetric surveys that are being suggested
by the SKA Magnetism and Continuum Science Working Groups for AGN at
SKA1-MID and SKA1-SUR, mainly on Bands 2 and 3 \citep{Tay14, Joh14,
  Pra14, Smo14}, will allow further investigations regarding the total
and polarized synchrotron emission of AGN.  While a measurement of the
spectra help to disentangle the contributions of AGN and star
formation and provide information about the AGN environment, the
measurement of the polarized fractions allows to further investigate
the anti-correlation between radio flux and polarization fraction, and
to potentially relate it to the power of the central engine, reflected
in the jet.  Whereas previous studies have found no explicit redshift
dependence in the observed correlation up to $z\sim3$, the far deeper
SKA1 surveys are expected to probe radio fluxes to $z\sim10$ where
correlations with redshift cannot be ruled out. 

On a second stage, deep follow-up SKA1-VLBI surveys of AGN will have
the capability to image much smaller scales to investigate the origin
of the I-p anti-correlation in terms of cancellation of orthogonal
polarized regions on bright sources.  These VLBI surveys will also
allow us for studies of the brighter spatially resolved jets on
cosmological scales, which will certainly help to provide information
on the environments (through VLBI morphological studies).  At L band,
a 5\,min-long VLBI observation including SKA1-MID will achieve a
$5\sigma$ level of 25\,$\mu$Jy.  This can allow to observe $\sim1000$
sources in less than 100 hours of observation.  The value of these
VLBI surveys will be much enhanced by synergies with deep optical
surveys such as SDSS \citep{Yor00}, and LSST \citep{Bac14}.

\section{Identifying AGN Populations from SKA1 Surveys}
\label{ident}
Two main processes contribute to the extragalactic continuum radio
emission at intermediate radio frequencies (i.e. 1.4 to 5 GHz): the
non-thermal emission associated with relativistic electrons powered by
AGN and a combination of free-free and synchrotron emission from the
star formation activity in galaxies.  The bright radio sky is
dominated by the emission driven by radio-loud AGN where the presence
of compact or slightly resolved radio cores on milliarcsecond--scale
are the best indicators of the AGN activity.  At faint flux
densities ($<$ 0.5 -- 1 mJy) the contribution from star-forming
galaxies to the radio source population becomes increasingly important
\citep[e.g.][and Fig.~\ref{pad}]{Seymour2008,Padovani_etal11}.
Moreover, recent work has confirmed that at 10--100 micro-Jy levels,
the radio-quiet AGN are not radio-silent \citep[e.g.][]{Bonzini12}.

Linear polarization and spectral index information from the massive
surveys on SKA1-MID and SKA1-SUR described above will be a useful tool
to disentangle AGN jet emission from that of star-forming galaxies,
thus providing a first selection of AGN candidates from large SKA
commensal surveys.  Clearly, the optical data and photometric
redshifts measured by the Large Synoptic Survey Telescope (LSST) will
be an invaluable complement for any kind of SKA survey involving
extragalactic sources, e.g. \citet{Bac14}, and specifically for all
those studies interested in radio emitting objects at all ranges of
redshift.  In particular, the optical photometric data from LSST will
also be extremely useful in discriminating between the AGN and star
forming galaxy populations.  A definitive probe of the presence of AGN
radio loud activity will be the detection of high brightness
temperature from embedded radio-AGN cores in the host galaxies through
ultra-high-resolution (milliarcsecond or sub-milliarcsecond scale)
SKA1-VLBI wide-field-of-view observations, see \citet{Par14}.

Adding to the present High Sensitive Array (HSA), the SKA1-MID and/or
SKA1-SUR as a phased array, will obtain an angular resolution better
than 5\,milliarcseconds at L Band, which corresponds to 10 -- 40 pc in
the redshift range 0.1 to 5.  This high angular resolution will be
coupled with the exceptional sensitivity provided by the SKA1.  One
hour of observation with the HSA + SKA1-MID at L Band at a recording
rate of 4096 Mb/s will allow to reach 0.75\,$\mu$Jy/beam
(0.84\,$\mu$Jy/beam without Arecibo).  For less demanding
sensitivities, similar VLBI observation will achieve a sensitivity of
$\sim5\,\mu$Jy/beam in 5\,min integrations.  This high sensitivity --
high angular resolution will allow not only identifying starburst
galaxies, and resolving and mapping the region of star-formation, but
also imaging jets at parsec scales in the inner region of AGN up to high $z$.

\section{Studies of Relativistic Jets in AGN with SKA1: A Summary}
Previous sections outline a variety of relevant science cases for
cutting-edge studies of relativistic jets in AGN that can be performed
with SKA1.  In particular, the superb sensitivity of SKA1 will allow
us to detect massive samples of AGN from commensal full-polarization
all-sky and deep surveys performed at both SKA1-MID and SKA1-SUR at
frequencies between $\sim1$ and 3\,GHz \citep{Tay14, Joh14, Pra14,
  Smo14}.  Identification of AGN in these surveys may be performed in
different and complementary ways by using classification arguments
based on linear polarization, spectral index, optical photometry, and
brightness temperature measurements.  Linear polarization and spectral
index information will be available from SKA commensal surveys.
Optical photometric and redshift data from the LSST will be of
invaluable help, whereas the highly-constraining
brightness-temperature information will be provided by follow up
wide-field VLBI observations involving SKA1, ideally including both
the phased SKA1-MID and SKA1-SUR at L-band, where UV coverage can be
optimized by including a large number of non-SKA VLBI stations.  The
wide field requirement needs multiple ($\ge4$) beam formers for the
SKA1-MID and SKA1-SUR cores, see \citet{Par14}.  These cores will
dominate the sensitivity of the entire VLBI array, and can therefore
reliably detect high brightness temperatures from AGN at a few
10\,$\mu\rm{Jy}$ level or better.

The data from the above-mentioned full-polarimetric surveys at
SKA1-MID and SKA1-SUR will be of almost inestimable value for
essentially all AGN science cases described in this chapter, specially
for radio-loud/radio-weak, and massive cosmological
statistical-studies.  Investigations of the feedback between large
scale AGN jets and their intergalactic medium, and those for the
search of the first GPS-like radio AGN and black holes, will also
benefit from these enormous survey datasets, as well as from lower
frequency all-sky programs at SKA-LOW.

The non-isotropic character of radio AGN, together with the fact that
some of the most pressing open questions about AGN are related to
their innermost (pc and sub-pc) jet regions, makes ultra-high angular
resolution VLBI observations an essential ingredient of
next-generation studies.  In particular, jet formation, collimation,
and acceleration studies will only be complete with the aid of
additional super-sensitive full-polarization VLBI observations at
multiple observing frequencies up to $\sim15$\,GHz, which will only be
possible with SKA1-MID on Band 5 and lower.  These observations are
also mandatory for detailed studies of the velocity field and magnetic
field properties of the innermost AGN jet regions, as well as for
research on the high energy emission site and emission mechanism in
combination with other multi-spectral range facilities operating
simultaneously in the southern hemisphere.

Characterizing the jet composition and magnetic field, and measuring
the low-energy cut-off of the relativistic electron energy
distribution through circular polarization analysis requires not only
multi-frequency (up to Band 5) SKA1-MID, and SKA1-VLBI observations to
map the different regions of the AGN, but also --and even more
importantly-- requires high precision circular and linear polarization
measurements up to $\sim0.01$\%, and $\sim0.1$\,\%, respectively, that
can be achieved with linear feeds as planned for SKA1-MID.  Great
attention must be paid at design time to ensure optimum performance in
all four Stokes parameters so that the scientific potential of SKA can
be realized in these areas.

\section{Considerations for SKA1 Early Science and SKA2}
A reduction by $\sim50$\,\% in sensitivity during the early-science
stage of SKA1 will not dramatically affect most of the science use
cases presented here as long as SKA1 receivers retain adequate
polarization purity as described above.  In particular, the proposed
research on jet formation, collimation and acceleration; jet
composition, precise characterization of jet parameters and jet
emission properties; jet feedback; and the identification of AGN
populations from early-science SKA surveys will provide excellent
science outcomes even with a large reduction in sensitivity while
waiting for the complete SKA1 to be finished.

The requirements for detailed imaging studies of FRII radio galaxies
outlined in \S~\ref{jetpar}, i.e. $\sim0.01$ arcsec resolution,
$\sim10$ nJy/beam, and $> 10^7:1$ dynamic range at bands 4 and 5 of
SKA1-MID, will only be achieved by SKA2.  Certainly, the
ability of SKA-MID to achieve frequencies up to 24\,GHz at the stage
of the SKA2 will greatly boost ultra-high-resolution VLBI studies
to probe the innermost jet regions of nearby sources, where the jet is
still being collimated and accelerated.
 
During the final stage of SKA, the possibility of SKA2-MID
observations including a number of very large ($\apgt1000$\,km)
baselines employing a homogeneous array of SKA stations would open a
new era of extremely high-dynamic-range VLBI observations with the SKA
alone \citep{Gar99}.  These high dynamic-range observations would
allow imaging of an unprecedented number of jet-counterjet systems,
and their transverse structure, which will certainly allow for
extremely high precision estimates of the actual physical conditions
at the innermost regions of jets, close to their formation and
collimation sites.  Of paramount importance for these SKA2-VLBI
observations is the availability of quasi-scaled arrays over as wide a
range of frequencies as possible to avoid changing resolution with
frequency, which is crucial to eliminate resolution biases on the
relevant polarization studies described in \S~2, 3, 4, and 6.



\end{document}